\documentclass[12pt]{article}
\usepackage{graphicx}
\usepackage{geometry}
\begin{document}
\setcounter{page}{0}
\thispagestyle{empty}
\renewcommand{\thesection}{\Roman{section}}
\begin{center}
\large{\bf STUDY OF SOME ASPECTS OF ASTRONUCLEAR SYSTEM}\\
\vspace{0.5cm}
\normalsize\rm
{Shamim Haque Mondal$ ^{1}$; Mahamadun Hasan$^{2}$; Murshid Alam$^{3}$; Md. Abdul Khan$ ^{*4} $}\\
\footnotesize
Department of Physics, Aliah University, \\ IIA/27, Newtown, Kolkata-700160, India\\ 
\vspace*{.5cm}
{\it Email:} $^1$shamimmondal709@gmail.com;
$^2$mahamadun.hasan@gmail.com, $^3$alammurshid2011@gmail.com, $^{4*}$drakhan.rsm.phys@gmail.com(*Corresponding author) \\
\end{center}
\vspace{1.0cm}
\rm
\begin{abstract}
In this work we have studied the bulk properties of compact stellar object which is largely motivated by the recent work of Potekhin group who studied the structure and evolution of compact star on the basis of unified EoS of Brussels-Montreal group. Here we solved ToV equation of state numerically using three sets of parameter models to compute pressure, baryon density and bulk modulus etc over a wide range of matter density. The computed values of the observables like the stellar interior pressure, bulk modulus and baryon density have been presented as a function of mass density in the tabular form. Graphical analysis of computed observables and their comparison with reference data wherever available have been presented as representative cases.
\end{abstract}

\hspace{1.0cm}{\it Keywords:} \\ \hspace*{1.0cm}\\ \hspace*{1.0cm}{\it PACS:} 

\section{Introduction}
Nuclear astrophysics is one of the emerging branches of physics which deals with physical phenomenons in happening in astronuclear systems. An astronuclear system is a natural physical system in which the long-range as well as the short-range interactions do play key roles. Innumerous phenomena occurring in the astronuclear systems provide ample evidence that matter exists in their extreme or critical condition in the astronuclear systems. For the prediction of a meaningful stellar model and for fruitful investigation of physical properties like transport phenomena, specific heat, entropy, Bulk modulus, incompressibility, etc of stellar matter, pressure-density relation which represents the equation of state (EOS) needs to be solved numerically subject to appropriate conditions. Compact stellar objects offer unique cosmic laboratories where matter at very high densities can be investigated. The Mass-radius (M-R) relation of such compact stellar objects can be obtained by solving TOV (Tolman Oppenheimer-Volkoff) equations\cite{tov} subject to appropriate matching conditions.

Literature review revealed the fact that there were some controversies regarding the compressibility (or compression modulus) of nuclear matter at the saturation limit \cite{glendenning}, and occurrence of giant monopole resonance in nuclei around 220 MeV \cite{blaizot} is accepted as the source of such properties, although Brown and Osnes pointed a relatively smaller energy $\sim 100$ MeV\cite{brown}. 

In the low energy limit, nuclear fusion reactions of light nuclei play a vital role in the primordial nucleosynthesis\cite{khn}. As the temperature and density increases, the CNO cycle becomes dominant till iron (Fe) is produced. The relics (or debris) of a star are not able to generate sufficient thermal pressure to resist against gravitational collapse, as a result, it undergoes appreciable contraction to reach an extremely high density. At this extreme density condition, nonthermal pressure is produced due to fermionic degeneracy and inter-fermion (nucleon, electrons, strange fermions, etc) interactions. Since, matter in compact objects, spans an enormous range of densities starting from few gm$/cm^{3}$ (for $_{26}^{56}$Fe) at the surface of a compact object to nucleonic matter density ($\rho \sim 10^{15}$ gm$/cm^{3}$) in the central region. Thus, weak-nuclear, strong-nuclear and the electromagnetic interactions become key players in the investigation of the structural properties of compact stellar objects.
One of the most studied most interesting compact objects is the neutron stars which are the densest and smallest stars other than black holes. They are born when a massive star ($1.4 M_{\odot}$ to $2 M_{\odot}$\cite{seeds}) runs out of fuel and collapses. Once the born, they no longer produce heat and continue to cool over time and chances are there that they may evolve further through collision or accretion. Most of the existing models of neutron stars imply that they are mostly composed of neutrons because at the neutron star condition electrons and protons combine to form neutrons. As neutron degeneracy pressure arising due to the Pauli exclusion principle, can prevent collapse to the extent of $0.7M\odot$\cite{tolman}\cite{oppenheimer}, the repulsive nuclear forces play a leading role in supporting more massive neutron stars \cite{douchin}. If the remnant star has a mass exceeding the Tolman–Oppenheimer–Volkoff limit of around $2M\odot$, the combination of degeneracy pressure and nuclear forces is insufficient to support the neutron star and it continues collapsing to form a black hole.

Neutron star being a compact stellar object incorporates a wide variety of physical phenomena including those from astro-nuclear physics, astroparticle physics, densed matter physics, and such others. At this stage of game it may be useful to mention that the predicted density of a neutron star varies aprroximately five order of magnitude when one makes a journey from its central region to its superficial region. Hence, a unified equation of state (EOS) suitable for the entire region of a neutron star matter will be an essential tool for extraction of information on the anatomy of a neutron star. The main objectives of the present study are two-fold. The first one is to focus on searching for an appropriate set of coupled nonlinear equations (CNE) which can elegantly connect the physical observables in the EOS. And the second one is to set up an effective model based on available data to investigate the bulk properties like the conductivity both thermal and electric, compressibility, superfluidity, etc of a neutron star. 

So far a significant number of papers have been published in the literature which mainly focus on the M-R relation of compact objects. Literature survey further reveals that the radii of compact objects are primarily determined by the pressure-density relation of nuclear matter under equilibrated condition. Under such condition, the radius is virtually independent of the mass and primarily determined by the magnitude of the pressure. And the pressure, at densities of the order of nuclear matter density practically becomes a function of the nuclear symmetry energy. The sun at its normal condition ($M_{\odot}\sim 10^{30}$kg, R$_{\odot}\sim 7\times 10^5$km) has an average matter density of $1.41 gm/cm^{3}$ with a central density of approximately $162.2 gm/cm^3$ \cite{sfs}. When this normal sun is squeezed into a ball of merely 10km radius its density reaches the nuclear matter density ($\sim 5\times 10^{14}gm/cm^3$) i.e., in such a situation the sun transforms to a super nuclear ball. If the mass of this ball is increased to approximately double the solar mass keeping the radius unaltered, the asymmetry energy starts dominating over the symmetry energy of the nuclear matter and the star starts missing its nuclear structure as a result of the capture of electrons by protons to give birth to new neutrons. So the matter inside the star practically transforms to neutron matter and hence a neutron star comes into existence. It is to be bourn in mind that the ordinary mass density relation (i.e., density = mass/volume) no longer applies to the dense matter, under that situation there exists a nonlinear logarithmic relation between mass density which is practically governed by the intervening pressure. 

Although there is significant progress in neutron-star physics as indicated by the discovery of the massive stars $\sim 2M_\odot$ \cite{Fonseca}\cite{Demorest}\cite{Antoniadis}, or with the discovery of neutron stars-merger event the GW170817 detected by LIGO and Virgo \cite{Abbott1}\cite{Abbott2}. In spite of the multidimensional advancements in neutron-star physics, there is still a lack of available data in the literature on the bulk properties of compact objects, and more specifically on the neutron stars. In this communication, we shall restrict ourselves focusing on the bulk properties of the observed neutron stars. 

This preliminary work is largely motivated by the recent work of Potekhin et al. (2013)\cite{potk}, where the authors have derived two EOS corresponding to two unified EOSs known as FPS EOS of Pandharipande and Ravenhall(1989)\cite{Raven} and SLy EOS of Douchin and Haensel (2001)\cite{Douch} with data given in the form of Tables, which were interpolated or extrapolated to meet the purposes.
In section II we briefly discuss the theoretical scheme adopted in our work, Section III will be devoted to results and discussion and finally, we will put our concluding remarks in section IV.


\section{Theoretical Scheme}
Generally, an EoS is formulated in tabular form containing baryon number density $n_{b} $, nuclear matter density $\rho $, and pressure P\cite{Haensel}. For cold catalyzed nuclear matter (i.e., matter fully equilibrited at $T=0$K) the Eos is determined by the functional dependence of the energy per baryon ($\epsilon$) on the baryon number density ($n_b$) i.e., $\epsilon = \epsilon(n_{b})$. The stellar mass density $\rho$ is connected to the rest energy density $\epsilon$ through the relation
\begin{equation}
\rho= \frac{\epsilon}{c^{2}}
\end{equation}
The pressure, P can be calculated from energy density $\epsilon$ by the application of the first law of thermodynamics in the limit of absolute zero temperature i.e., T = 0 limit.
\begin{equation}
P(n_{b})=n_{b}^2\frac{d}{dn_{b}}\epsilon(n_{b})=n_{b}^{2}c^{2}\frac{d}{dn_{b}}(\frac{\rho}{n_{b}})
\end{equation}
The adiabatic index ($\Gamma$), which measures the degree of stiffness of the EoS at given density is defined as 
\begin{equation}
\Gamma=\frac{d (\log P)}{d (\log n_{b})}
\end{equation}
It is an interesting dimensionless paramater since different regions of stellar matter interior are chaecterized by distinct behaviour of $\Gamma$. 

Douchin and Haensel(2001) \cite{Douch} have studied stiffness of stellar matter in terms of $\Gamma$. Their results indicated a significant increase in the stiffness parameter at the crust-core interface. Where the values of $\Gamma $ drastically changes from 1.7 to 2.2 \cite{Phaen}. Using Eqs. (1), (2) \& (3) we can get an estimate of the acoustic wave velocity in the stellar medium in terms of Newtonian-Laplace formula
\begin{equation}
v_s = \sqrt{\frac{\Gamma P}{\rho}}
\end{equation} 

The above relations (Eqs. (1) and (2)) may also be used to study the bulk modulus (B) as a function of the baryon density and can be expressed as 
\begin{equation}
B(n_{b})=n_{b}\frac{dP}{dn_{b}}=[1+\frac{P}{\rho c^{2}}]\frac{\rho}{P}\frac{dp}{d \rho}
\end{equation}
Potekhin et al.(2013)\cite{potk} presented numerical values of pressure (P), gravitational mass density ($\rho$) as function of the baryon number density ($n_b$). The authors also demonstrated the applicability of their results to investigate basic characteristics of cold dense matter such as that for a neutron star. They studied the mass-radius (M-R) relations, limiting (i.e., maximum and minimum) masses, thresholds of direct Urca processes, electron conductivity, etc in the neutron-star crust, in the adopting different EoS models namely BSk19, BSk20, and BSk21 based on their analytical representations in particular with different stiffness parameters. It is worth mentioning here that the work of Potekhin group is based on the Brussels-Montreal unified EOS derived in terms of energy-density functional theory (EDF)\cite{Gror},\cite{Cham}. The so-called EoS models BSk19, BSk20 and BSk21 corresponds to different stiffness factor. With the help of the simulations of Potekhin et al, we computed bulk modulus (B) and baryon density ($n_{b}$) over a wide range of nuclear matter densities (i.e., sub-nuclear, nuclear and supranuclear) of neutron star matter according to relation (4).

\section{{Results and Discussion}}
Calculated baryon density($n_b$), pressure(P), bulk modulus(B), sound velocity(vs) for gradually increasing mass density are presented in columns 2, 4, 6 and 7 of Table 1 respectively, corresponding reference values for baryon density, pressure, and sound velocity are also presented in columns 3, 5 and 8 respectively. In the present work, we have expressed bulk modulus (B) and pressure (P) in mega-Newtown per fermi$^2$ (i.e., MN fm$^-2$) unit.
Computed values of the bulk modulus (B) for different nuclear mass density ($\rho$) is plotted in logarithmic scale as shown in Figure 1. The reference data are taken from Baym et al (1971) \cite{Bps}, Pandharipande et al (1998)\cite{Pnd}, also plotted on the same graph, show an excellent agreement with our computed data. Table 1 contains only representative data points, the graph is plotted using a very large set of data points computed for matter densities ranging from $10^6 g\:cm^{-3}$ to $10^{16} g\:cm^{-3}$.

\begin{figure}
\centering
\fbox{\includegraphics[width=0.6\linewidth, height=0.4\linewidth]{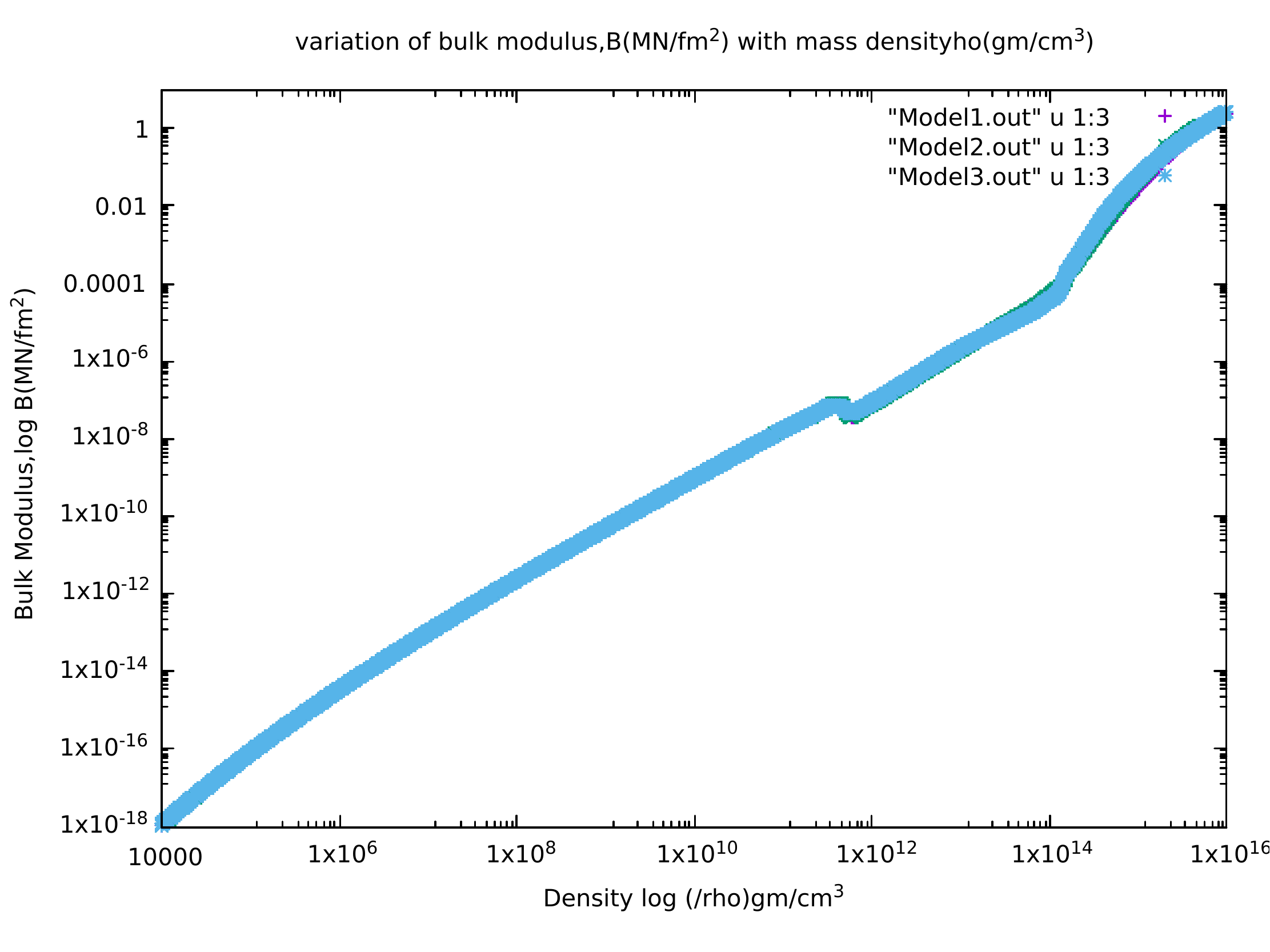}}
\caption{Variation of bulk modulus, B with mass density, $\rho$ for different EoS Model}
\label{fig:boxed_graphic}
\end{figure}

As can be seen from Figure 1, the bulk modulus (B) changes smoothly with mass density ($\rho$) up to about $\rho \simeq 10^{16} g\:cm^{-3}$. Around $\rho \simeq 5\times 10^{12} g\:cm^{-3}$ there is a kink in the graph indicating a possible phase change in the nuclear matter which is also reflected in the very very recent works of Baym \cite{baym}. This may be due to the change in phase of stellar matter arising due to a rapid conversion of 1st generation quarks which made up the hadron components into strange quarks giving rise to the increase in fractional density of strange particles. This whole process facilitates the stability of neutron star matter up to $2M\odot$. And thereafter new phase arises which needs further investigation. Thereafter, the step-like graph in the density range $10^{12} g\:cm^{-3}$ to $10^{14} g\:cm^{-3}$ possibly corresponds to the dripline regime for the first generation quark matter. Beyond $\rho \sim 10^{14} g\:cm^{-3}$ the slope of the graph becomes relatively steeper which corresponds to the later generation quark matter composition.
The standard BPS EoS\cite{Bps} suggests a neutron drip density, $\rho_{n-drip} (\sim 4\times 10^{11} g\: cm^{-3})$. Mass density, $\rho \leq \rho_{n-drip}$ characterizes matter in the outermost layers of neutron stars where as the same for $\rho > \rho_{n-drip}$ describes the matter in the interior of the neutron stars. 

\begin{table}
\begin{center}
\begin{scriptsize}
{\bf Table 1. Equation of state of cold, catalyzed high density matter for model 3}\\
\vspace{6pt}
\begin{tabular}{l|ll|ll|c|ll}\hline\hline
Mass density&\multicolumn{2}{c}{Barion density, $n_{b}$($fm^{-3}$)}\vline&\multicolumn{2}{c}{Pressure, P$(MNfm^{-2})$}\vline&\ Bulk Modulus(B)&\multicolumn{2}{c}{Velocity, $V_{s}$(in unit of c)}\\
\cline{2-3}\cline{4-5}\cline{7-8}
$ \rho (gm cm^{-3})$& Calculated&Reference&Calculated&Reference&Calculated&calculated&Reference\\
&&&&& $(MNfm^{-2})$&&\\\hline
1.00E+06&6.01E-10&6.02E-10&2.17E-15&3.63E-15&3.44E-15&0.0057c&0.0063c\\
3.00E+06&1.81E-09& &1.19E-14& &1.80E-14&0.0076c& \\
9.00E+06&5.49E-08& &6.85E-14& &8.80E-14&0.0106c& \\
1.00E+07&6.02E-09&6.02E-09&7.05E-14&6.81E-14&1.02E-13&0.0102c&0.0107c\\
5.00E+07&3.02E-08&3.01E-08&6.86E-13&6.84E-13&9.48E-13&0.0143c&0.0146c\\
1.00E+08&6.02E-08&6.02E-08&1.77E-12&1.78E-12&2.41E-12&0.0162c&0.0169c\\
6.00E+08&3.64E-06& &1.94E-11& &2.56E-11&0.0219c& \\
1.00E+09&6.02E-07&6.02E-07&3.80E-11&3.91E-11&4.97E-11&0.0237c&0.0225c\\
5.00E+09&3.06E-06&3.01E-06&3.06E-10&3.03E-10&3.94E-10&0.0101c&0.0101c\\
1.00E+10&6.01E-06&6.01E-06&7.46E-10&7.24E-10&9.56E-10&0.0332c&0.0337c\\
5.00E+10&2.99E-05&3.00E-05&5.78E-09&5.63E-09&7.28E-09&0.0414c&0.0377c\\
1.00E+11&6.01E-05&5.99E-05&1.37E-08&1.40E-08&1.68E-08&0.113c&0.0447c\\
5.00E+11&2.99E-04&2.99E-04&8.61E-08&8.28E-08&5.44E-08&0.0503c&0.0321c\\
1.00E+12&5.99E-04&5.97E-04&1.27E-07&1.26E-07&8.58E-08&0.0434c&0.034c\\
5.00E+12&2.99E-03&2.98E-03&6.45E-07&4.79E-07&8.26E-07&0.0817c&0.0344c\\
6.00E+12&3.58E-03& &8.17E-07& &1.07E-06&0.045c& \\
1.00E+13&5.95E-03&5.95E-03&1.61E-06&1.16E-06&2.13E-06&0.0487c&0.0444c\\
5.00E+13&2.96E-02&2.97E-02&1.18E-05&1.32E-05&1.35E-05&0.059c&0.067c\\
1.00E+14&5.91E-02&5.92E-02&2.78E-05&3.93E-05&3.83E-05&0.064c&0.096c\\
6.00E+14&3.41E-01& &5.12E-03& &1.63E-02&0.3567c& \\
1.00E+15&5.32E-01&5.50E-01&1.96E-02&1.52E-02&5.63E-02&0.5367c&0.7667c\\
1.00E+16&1.32E+01& &9.89E-01& &2.72E+0&$\simeq c$& \\ \hline\hline
\end{tabular}
\end{scriptsize}
\end{center}
\end{table}

\begin{figure}
\centering
\fbox{\includegraphics[width=0.6\linewidth, height=0.6\linewidth]{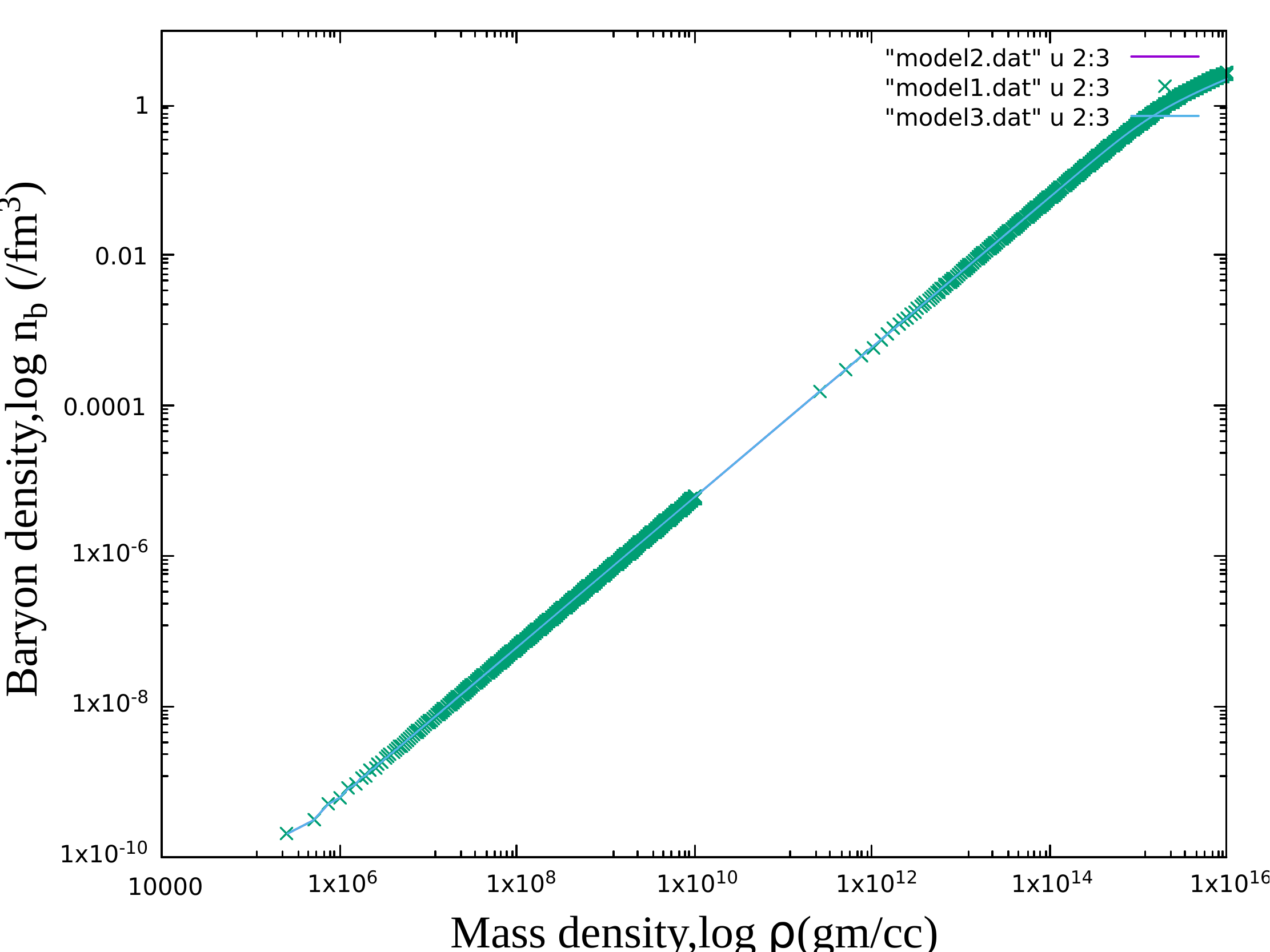}}
\caption{Variation of Baryon density, $n_{b}$ with mass density, $\rho$ for different EoS Model}
\label{fig:boxed_graphic}
\end{figure}
\begin{figure}
\centering
\fbox{\includegraphics[width=0.6\linewidth, height=0.4\linewidth]{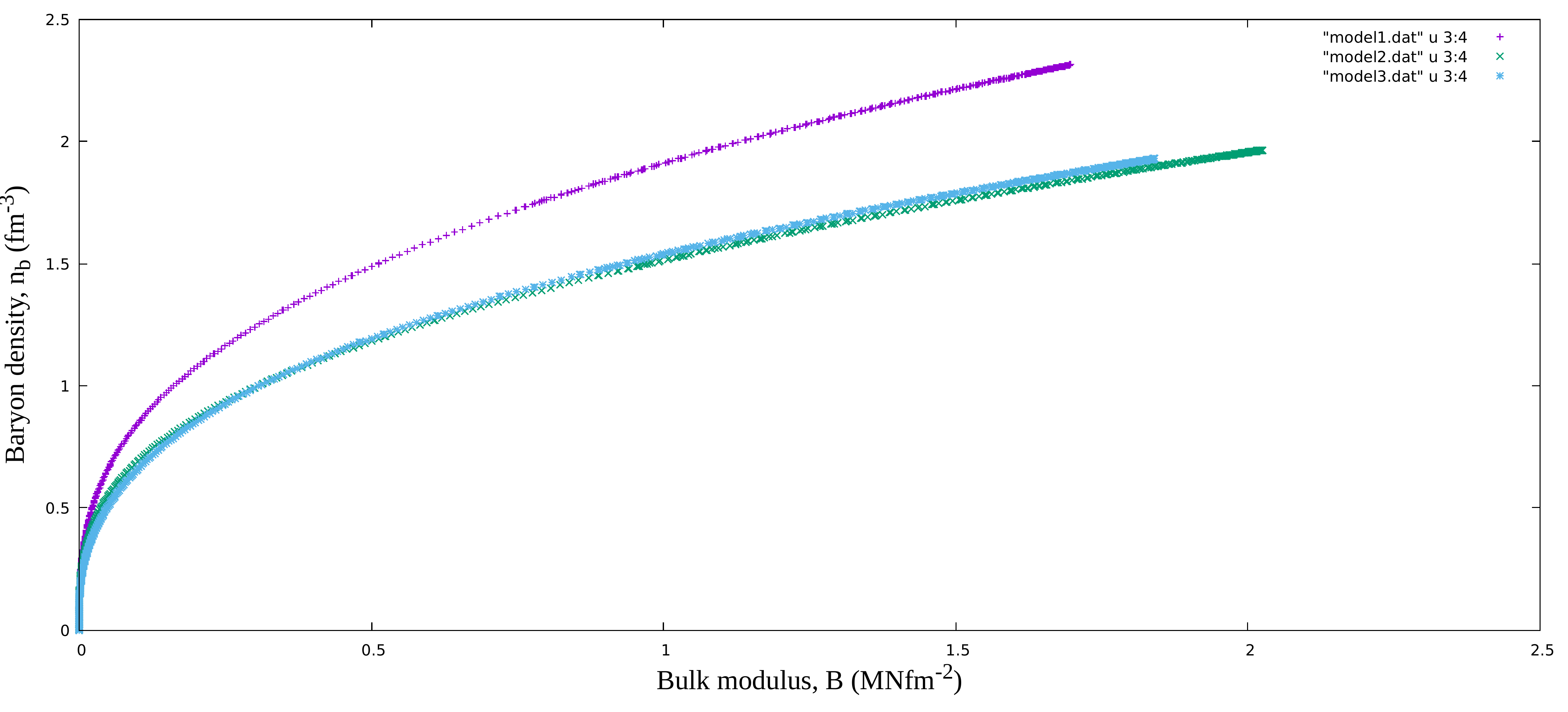}}
\caption{Variation of bulk modulus, $\log$ B with baryon density, $\log n_{b}$ for different EoS Model}
\label{fig:boxed_graphic}
\end{figure}

Figure 2 has been obtained by plotting the baryon density $n_{b}$ as a function of the mass density $\rho$. As expected, the baryon density increases almost linearly with the mass density tending towards saturation value beyond $10^{15} g\: cm^{-3}$. Beyond the saturation limit, baryon density may decrease due to the decay of baryons and the corresponding increase in the light components like lepton, quarks, etc. We have also plotted the baryon density ($n_b$) as a function of bulk modulus (B) as depicted in Figure 3. The plot indicates a requirement of a relatively greater baryon density population growth rate in the lower stiffness end than in the larger stiffness region. 

\section{{Conclusions}}
Incompressibility is more commonly used in nuclear physics applications rather than compressibility ($\chi$) which is the reciprocal of Bulk modulus, B (i.e., $ \chi= B^{-1} $). So the study of B gives a better picture of the nature of the stellar matter which most of the times exhibit exotic behavior. It may be noted that the computed sound velocity ($v_s$, expressed in units of light speed in free space) values presented in column 7 of Table I exhibit random fluctuation which may be crept in due to randomness in the adiabatic index($\Gamma$) involved in the calculation. Neutron Star the value of B is $10^{23}$ times higher than the value of the same for the matter at the interior of the sun. So it is intuitively expected that the properties of stellar materials are very much different from terrestrial material. For example, the speed of the acoustic wave (e.g., sound) in a neutron star is expected to be several tens of percent of the speed of light!\cite{Shapir}. Finally, it may be stated that an in-depth study of the bulk properties of dense stellar matter may yield very important information even the effect of the gravitational wave can be inferred from the study subject to appropriate parametrization.

The authors acknowledge Aliah University for providing the computational facility. One of the authors S H Modal acknowledges the Govt. of West Bengal, India for financial assistance in the form of SWVMS fellowship.


\end{document}